\documentclass[twocolumn,floats,prb,amsmath,showpacs,superscriptaddress]{revtex4}

\usepackage{epsfig}
\usepackage{graphicx}
\usepackage[usenames,dvipsnames]{color}

\begin{document}

\newcommand{\ket}    [1]{{|#1\rangle}}
\newcommand{\bra}    [1]{{\langle#1|}}
\newcommand{\braket} [2]{{\langle#1|#2\rangle}}
\newcommand{\bracket}[3]{{\langle#1|#2|#3\rangle}}

\def\bea{\begin{eqnarray}} 
\def\nn{\nonumber\\}
\def\eea{\end{eqnarray}}
\def\beq{\begin{equation}}
\def\eeq{\end{equation}}

\def\fac{\gamma}
\def\facc{\gamma_c}
\def\facbar{\overline\fac}
\def\faccbar{{\overline\fac}_c}

\def\one{\hat{\bf 1}}
\def\one{\hat{\it 1}}
\def\pw{^{({\rm W})}}
\def\ph{^{({\rm H})}}
\def\k{{\bf k}}
\def\r{{\bf r}}
\def\R{{\bf R}}
\def\0{{\bf 0}}
\def\b{{\bf b}}
\def\q{{\bf q}}
\def\o{{\cal O}}
\def\e{{\cal E}}
\def\v{{\bf v}}
\def\c{c^{\dagger}}
\def\m{{\bf m}}
\def\M{{\bf M}}
\def\Morb{\M_{\rm orb}}
\def\x{\hat x}
\def\y{\hat y}
\def\P{\hat P}
\def\Q{\hat Q}
\def\H{\hat H}
\def\sr{\M_{\rm SR}}
\def\srone{\M_{\rm SR}^{(\rm I)}}
\def\sronez{M_{{\rm SR},z}^{(\rm I)}}
\def\srtwo{\M_{\rm SR}^{(\rm II)}}
\def\ic{\M_{\rm IC}}
\def\icint{\M_{\rm IC}^{({\rm int})}}
\def\icsurf{\M_{\rm IC}^{({\rm surf})}}
\def\deltaM{\boldsymbol\Delta\M}
\def\ppartial{\widetilde\partial}

\newcommand{\red}{\textcolor{red}}
\newcommand{\blue}{\textcolor{blue}}
\newcommand{\green}{\textcolor{OliveGreen}}
\newcommand{\cyan}{\textcolor{cyan}}
\newcommand{\magenta}{\textcolor{magenta}}

\newcommand{\ivo}[1]{\marginpar{\red{\textsf{[\small ivo: #1]}}}}
\newcommand{\dvm}[1]{\marginpar{\green{\textsf{[\small dhv: #1]}}}}
\newcommand{\ok}{\green{\textsf{ OK.}}}

\title{
Dichroic $f$-sum rule and the orbital magnetization of crystals
}

\author{Ivo Souza}
\affiliation{Department of Physics, University of California,
Berkeley, CA 94720, USA}

\author{David Vanderbilt}
\affiliation{Department of Physics and Astronomy, Rutgers University,
Piscataway, New Jersey 08854-8019, USA}

\marginparwidth 1.0in
\marginparsep 2.5in

\date{\today}
\begin{abstract}
We consider the magnetic circular dichroism spectrum of a crystal with broken
time-reversal symmetry in
the electric-dipole approximation. Using the Kubo-Greenwood formula for the
absorptive part of the antisymmetric optical conductivity, 
its frequency integral 
is recast as a ground-state property. We show that in insulators
this quantity 
is proportional to the circulation of the occupied Wannier orbitals
around their centers (more precisely, to the gauge-invariant part thereof).
This differs from the net circulation, or ground state
orbital magnetization, which has two additional contributions: (i) the
remaining Wannier self-rotation, and
(ii) the ``itinerant'' circulation arising from the
center-of-mass motion of the Wannier orbitals, both on the surface and
in the interior of the sample. Contributions (i) and (ii) are not
separately meaningful, since their individual values depend on the
particular choice of Wannier functions. Their sum is however
gauge-invariant, and  can be inferred from a
combination of two experiments: a measurement of the magneto-optical spectrum
over a sufficiently wide range to evaluate the sum rule, and a
gyromagnetic determination of the total orbital magnetization.

\end{abstract}
\vskip 2mm

\maketitle
\vskip2pc
\marginparwidth 3.1in
\marginparsep 0.5in 

\section{Introduction}
\label{sec:intro}

Optical sum rules provide a link between excitation spectra and
ground-state properties. The best-known example is the $f$-sum rule of atomic
physics.\cite{sakurai} 
It relates the frequency-integrated
absorption of linearly polarized light to the number of valence electrons.
In this work we consider the analogous result for circularly
polarized light. For non-magnetic systems the circular $f$-sum rule
is simply the average
of the $f$-sum rules for the two linearly-polarized components of
the beam, again yielding the total number of electrons.
If, however, the system is
magnetized, either spontaneously or by an applied field, this is no longer
the case; there is a small correction that flips sign when either the
magnetization of the sample
or the helicity of the incident light is reversed.
We are interested in what information this correction to the circular
$f$-sum rule provides about the magnetization.

The differential absorption of right- and left-circularly-polarized light
by magnetic materials is known as magnetic circular dichroism (MCD). The
object of interest in this work can thus be viewed as a
``dichroic'' $f$-sum rule for the integrated MCD spectrum. Such a
sum rule was first derived by Hasegawa and Howard for the special case of
a hydrogen atom in a magnetic field.\cite{hasegawa61} They showed that it
is proportional to the quantum-mechanical expectation value of the
orbital angular momentum operator, i.e., to the orbital moment.
It has been assumed that this conclusion generalizes trivially
to many-electron systems such as solids.\cite{smith76,thole92}
This is not so,\cite{explan-sumrules} as shown by Oppeneer, who
obtained the correct sum rule for
that case.\cite{oppeneer98} He observed that it yields a
quantity that is subtly different from the orbital 
magnetization $\Morb$, and should
instead be viewed as one of two terms adding up to $\Morb$.

In a separate development, a rigorous theory of orbital magnetization in
crystals was recently 
formulated.\cite{timo05,xiao05,ceresoli06,shi07}
Interestingly, it also identifies two
separate contributions to $\Morb$. One key result of
the present work is to recast the dichroic $f$-sum rule
in the language of this modern theory, elucidating its physical
content. Conversely, the sum rule solves an open problem in the theory
of Refs.~\onlinecite{timo05,xiao05,ceresoli06,shi07} as raised
explicitly in Ref.~\onlinecite{ceresoli06}: whether the
two gauge-invariant contributions to $\M_{\rm orb}$ identified
therein are separately measurable in principle.  The present work
answers this question in the affirmative.

Although we will mostly focus on crystalline solids, we find it useful
to start in Sec.~\ref{sec:bounded} by discussing the sum rule in the more
general context
of bounded samples under open boundary conditions. The detailed treatment of
periodic crystals is deferred until Sec.~\ref{sec:bulk}.
In both cases, special emphasis will be placed on insulating systems, 
for which an intuitive real-space picture in terms of occupied Wannier 
orbitals can be given.
We conclude in Sec.~\ref{sec:summary} with a summary and outlook.
In Appendices~\ref{app:other}, \ref{app:many_body}, and \ref{app:thm_limit}
we derive and elaborate on some results quoted in the main text. In
particular, Appendix~\ref{app:other} discusses the relation between the
dichroic $f$-sum rule and three other known sum rules. 

\section{Bounded samples}
\label{sec:bounded}

\subsection{Preliminaries}

In this work we are interested in systems displaying broken time-reversal
symmetry in the spatial wavefunctions. A typical example would be
a ferromagnet such as iron in which the exchange interaction breaks
time-reversal symmetry in the spin channel and this symmetry breaking
is then transmitted to the orbital degrees of freedom by the
spin-orbit interaction.
Other examples include systems in applied magnetic fields,
and also certain spinless model Hamiltonians such as the Haldane
model.\cite{haldane88}

We work in the independent-particle approximation.  The interaction
with light will be treated in the  electric-dipole approximation,
valid at not-too-high frequencies. This should be adequate provided
that the sum rule saturates before higher-order contributions, such
as  electric quadrupole  and magnetic dipole terms, become
significant.  The
oscillator strength for the transition between one-electron states $n$ and $m$
is 
\beq
f^{(\hat{\boldsymbol\epsilon})}_{n\rightarrow m}=
\frac{2m_e}{\hbar\omega_{mn}}
\left|\bra{m}\hat{\boldsymbol\epsilon}\cdot\hat \v\ket{n}\right|^2.
\eeq
This expression, valid for a general polarization
$\hat{\boldsymbol\epsilon}$ of light, can be derived
in the same way\cite{sakurai} as the familiar oscillator strength formula
for linear polarization
$\hat{\boldsymbol\epsilon}=\hat{\bf x}$.
For light propagating along $\hat{\bf z}$ with circular polarization
\beq
\hat{\boldsymbol\epsilon}_\pm=
\frac{\hat{\bf x}\pm i\hat{\bf y}}{\sqrt{2}}
\eeq
(``+'' corresponds to positive helicity, or left-circular polarization),
\beq
\label{eq:f_circ}
f^{(\pm)}_{n\rightarrow m}=\frac{1}{2}
\left[
  f'_{nm,xx}+
  f'_{nm,yy}
\right]
\mp f''_{nm,xy},
\eeq
where we have introduced the matrix
\beq
\label{eq:f}
f_{nm,\alpha\beta}=
\frac{2m_e}{\hbar\omega_{mn}}\bra{n}\hat v_\alpha\ket{m}
\bra{m}\hat v_\beta\ket{n}.
\eeq
Here $\alpha,\beta$ label Cartesian directions, $\hbar\omega_{mn}=\e_m-\e_n$,
and $m_e$ is the electron mass. Note that the matrix $f=f'+if''$
is Hermitian in the Cartesian indices. Thus its real and imaginary parts
are symmetric and antisymmetric respectively. 

According to Eq.~(\ref{eq:f_circ}), the sum 
of the oscillator strengths for
the two circular polarizations $\hat{\boldsymbol\epsilon}_+$ and
$\hat{\boldsymbol\epsilon}_-$ equals the sum of the oscillator strengths
for the two linear polarizations $\hat{\bf x}$ and $\hat{\bf y}$, and 
is related to $f'$.
The circular dichroism, i.e., the
difference between the two circular oscillator strengths, is
given by $f''$:
\beq
\label{eq:dichroism_bounded}
f''_{nm,xy}=\frac{1}{2}
\left[
  f^{(-)}_{n\rightarrow m}-f^{(+)}_{n\rightarrow m}
\right].
\eeq

Consider now a macroscopic system (e.g., a sample of volume $V$ cut from a
bulk crystal) and decompose its optical
conductivity $\sigma_{\alpha\beta}(\omega)$ in three different
ways: (i) real and imaginary parts, $\sigma'$ and $\sigma''$; (ii)
symmetric and antisymmetric parts, $\sigma_{\rm S}$ and $\sigma_{\rm A}$;
(iii) Hermitian and anti-Hermitian parts, $\sigma_{\rm H}$ and
$\sigma_{\rm AH}$. Then
\begin{equation}
\label{eq:sigma_h}
\sigma_{\rm H}=\sigma'_{\rm S}+i\sigma''_{\rm A}
\end{equation}
and
\begin{equation}
\label{eq:sigma_ah}
\sigma_{\rm AH}=\sigma'_{\rm A}+i\sigma''_{\rm S},
\end{equation}
where the Cartesian indices and the frequency have been omitted.
The properties of $\sigma$ can be summarized by noting that
the Hermitian part is dissipative while the anti-Hermitian part
is reactive, and the symmetric part is ``ordinary'' while the
antisymmetric part is ``dichroic.''
At $T=0$ the dissipative (or absorptive) part is
\begin{equation}
\label{eq:sigma}
\sigma_{\rm H}(\omega)=\frac{\pi e^2}{2m_eV}\sum_n^{\rm occ}\sum_m^{\rm empty}
f_{nm}\delta(\omega-\omega_{mn}),
\end{equation}
where $-e$ is the electron charge.
The analog of Eq.~(\ref{eq:f_circ}) in terms of conductivities is
\beq
\sigma^{(\pm)}_{{\rm abs}}=
\frac{1}{2}
\left[
  \sigma'_{{\rm S},xx}+\sigma'_{{\rm S},yy}
\right]
\mp \sigma''_{{\rm A},xy}.
\eeq
Thus the difference in absorption between light with negative and positive
helicity is given by twice the imaginary part of the antisymmetric optical
conductivity,\cite{bennett65}
\begin{equation}
\sigma''_{{\rm A},xy}=
\frac{1}{2}
\left[
  \sigma^{(-)}_{{\rm abs}}-\sigma^{(+)}_{{\rm abs}}
\right].
\end{equation}
Like other magneto-optical effects, MCD
vanishes for time-reversal-invariant systems. This can be seen from the
Onsager relation $\sigma_{\alpha\beta}({\bf H},{\bf M})=
\sigma_{\beta\alpha}(-{\bf H},-{\bf M})$,
which implies $\sigma_{{\rm A},\alpha\beta}({\bf H}={\bf M}=0)=0$.
\subsection{Dichroic $f$-sum rule}
\label{sec:mcd_sum_rule}

With the notation
\beq
\label{eq:bracket}
\langle f\rangle \equiv \int_0^\infty f(\omega)d\omega ,
\eeq
the dichroic $f$-sum rule relates the integrated MCD spectrum
$\langle\sigma''_{{\rm A},\alpha\beta}\rangle$ to a
certain ground-state property of the system.  To see how, we begin
by expressing $\sigma''_{{\rm A},\alpha\beta}(\omega)$ as the
imaginary part of the Kubo-Greenwood formula (\ref{eq:sigma}). Combining with
Eq.~(\ref{eq:f}) and taking the integral,
\beq
\langle\sigma''_{{\rm A},\alpha\beta}\rangle=
\frac{\pi e^2}{\hbar V}\sum_n^{\rm occ}\sum_m^{\rm empty}\,
{\rm Im}\left(
              \frac{\bra{n}\hat v_\alpha\ket{m}\bra{m}\hat v_\beta\ket{n}}
                   {\omega_{mn}}
        \right).
\eeq
Using the identity
\beq
\label{eq:vel_pos}
\frac{\bra{n}\hat v_\alpha\ket{m}}{\omega_{mn}}=
-i\bra{n}\hat r_\alpha\ket{m}
\eeq
and defining the projector onto the empty states
$\hat Q=\sum_m^{\rm empty}\ket{m}\bra{m}$,
\beq
\langle\sigma''_{{\rm A},\alpha\beta}\rangle=
-\frac{\pi e^2}{2\hbar V}\sum_n^{\rm occ}\bra{n}\hat r_\alpha
\hat Q\hat v_\beta\ket{n}-(\alpha\leftrightarrow\beta).
\eeq
Introducing the pseudo-vector
$\sigma''_{{\rm A},\gamma}=
(1/2)\epsilon_{\alpha\beta\gamma}\sigma''_{{\rm A},\alpha\beta}$
and $\hat P=\sum_m^{\rm occ}\ket{n}\bra{n}$,
this can be written more concisely as
\beq
\label{eq:mcd_sum_rule_bounded}
\langle\boldsymbol\sigma''_{\rm A}\rangle=
-\frac{\pi e^2}{2\hbar V}{\rm Tr}[\P\hat\r\times\Q\hat\v].
\eeq

Eq.~(\ref{eq:mcd_sum_rule_bounded})
is the dichroic $f$-sum rule, also obtained in Ref.~\onlinecite{oppeneer98}.
Using the closure relation
$\Q=\one-\P$, it becomes apparent that the right-hand-side depends
exclusively on the occupied states, and is closely related to the 
total (macroscopic) ground-state
orbital magnetization $\M_{\rm orb}=\fac{\rm Tr}[\P\hat\r\times\hat\v]$, where
$\fac=-(e/2cV)$ in electrostatic units (esu). Writing
\beq
\label{eq:M_decomp}
\M_{\rm orb}=\srone+\deltaM,
\eeq
with
\beq
\label{eq:srone}
\srone=\fac{\rm Tr}[\P\hat\r\times\Q\hat\v]
\eeq
and
\beq
\label{eq:deltaM}
\deltaM=\fac{\rm Tr}[\P\hat\r\times\P\hat\v]
\eeq
(the notation will be explained shortly), Eq.~(\ref{eq:mcd_sum_rule_bounded})
becomes
\beq
\label{eq:mcd_sum_rule_bounded_b}
\langle\boldsymbol\sigma''_{\rm A}\rangle=\frac{\pi ec}{\hbar}\srone.
\eeq
Hence the sum rule yields an orbital quantity $\srone$
with units of magnetization, but differing from the actual orbital
magnetization by the remainder $\deltaM$.

Two of the three quantities in Eq.~(\ref{eq:M_decomp}) are
independently measurable.  The left-hand side can be
determined from gyromagnetic experiments,\cite{kittel53,scott62,huguenin71} 
while $\srone$ on the right-hand side is obtainable from magneto-optical
experiments via the sum rule.  Thus, their difference $\deltaM$ can
also be determined in principle.  However, measuring
$\srone$ and $\deltaM$ independently will be of only limited
interest unless some physical meaning can be attached to
each of them separately. With this goal in mind we shall now make
contact with the recent theory of macroscopic orbital magnetization.

\subsection{Relation to the orbital magnetization}
\label{sec:relation_morb}

The results obtained so far are fairly general. To proceed further
we specialize to insulating samples.
For the present purposes ``insulating'' means that the
ground state wavefunction can be written as a Slater determinant of
well-localized orthonormal molecular orbitals $\ket{w_i}$, which we will
generically refer to as Wannier functions (WFs) even when the sample does not
have a crystalline interior.\cite{mv97}
This definition encompasses a broad range of systems, both macroscopic
and microscopic, but it excludes
metals and Chern insulators,\cite{timo06} which are not
Wannier-representable in the above sense.

By invariance of the trace, the orbital magnetization can be expressed in the
Wannier representation as
\beq
\Morb=\fac\sum_i^{\rm occ}\,\langle w_i|\hat\r\times\hat\v|
w_i\rangle.
\eeq
In Ref.~\onlinecite{timo05} this was decomposed as\cite{explan-notation}
\beq
\label{eq:sr_ic}
\Morb=\sr+\ic
\eeq
where
\beq
\label{eq:Msr}
\sr=\fac\sum_i^{\rm occ}\,\bra{w_i}(\hat\r-\overline\r_i)\times\hat\v\ket{w_i}
\eeq
arises from the circulation of the occupied WFs around their
centers $\overline\r_i=\langle w_i|\hat\r|w_i\rangle=\r_{ii}$
(``self-rotation''), while
\beq
\label{eq:Mic}
\ic=\sum_i^{\rm occ}\,\overline\r_i\times\bra{w_i}\hat\v\ket{w_i}
\eeq
is the circulation arising from the
motion of the centers of mass of the WFs.

It is well known that the WFs of a given system are not uniquely defined;
unitary mixing among the WFs is allowed, giving rise to a ``gauge
freedom'' 
(not to be confused with the freedom to choose the electromagnetic gauge).
In practice one deals with this
issue by choosing, among the infinitely many possible gauges, a particular one
that has certain desirable properties. A common strategy is to work in
the gauge
that minimizes the quadratic spread of the WFs, producing so-called
maximally-localized WFs.\cite{mv97} Naturally, any physical observable
(e.g., $\Morb$) is necessarily invariant under a change of gauge.
This is unfortunately not the case for the individual terms $\sr$ and $\ic$
in Eqs.~(\ref{eq:Msr})--(\ref{eq:Mic}), which turn out to be gauge-dependent.
This is to be expected since these quantities do not take the
form of traces, unlike those in the decomposition introduced earlier
via Eqs.~(\ref{eq:M_decomp})--(\ref{eq:deltaM}).

The two decompositions (\ref{eq:M_decomp})--(\ref{eq:deltaM}) and
(\ref{eq:sr_ic})--(\ref{eq:Mic}) are not unrelated, however. To see this,
we insert the identity $\one=\Q+\P$ at the location of the cross
product in Eq.~(\ref{eq:Msr}) to obtain
\beq
\label{eq:sr_one_two}
\sr=\srone+\srtwo,
\eeq
where $\srone$ is the quantity defined in Eq.~(\ref{eq:srone})
(since $\bra{w_i}\Q=0$), and
\bea
\label{eq:srtwo}
\srtwo&=&\fac
\Big(
{\rm Tr}[\P\hat\r\times\P\hat\v]-\sum_i^{\rm occ}\,
\overline{\r}{_i}\times\overline\v_i
\Big)\nn
&=&\fac\sum_{i,j\not=i}^{\rm occ}\,\r_{ij}\times\v_{ji}.
\eea
In this way we have segregated the gauge-dependence of $\sr$ to the term
$\srtwo$, isolating a gauge-invariant part $\srone$ which
turns out to be precisely the quantity defined in Eq.~(\ref{eq:srone})
and appearing in the sum rule (\ref{eq:mcd_sum_rule_bounded_b}).
When the gauge-dependent self-rotation $\srtwo$ is combined with the
gauge-dependent itinerant circulation $\ic$, it forms the
gauge-invariant quantity
$\deltaM$ of Eq.~(\ref{eq:deltaM}).  The relation between the
decompositions (\ref{eq:M_decomp})--(\ref{eq:deltaM}) and
(\ref{eq:sr_ic})--(\ref{eq:Mic}) can be summarized by writing
\beq
\label{eq:m_orb_decomp}
\M_{\rm orb}=\M_{\rm SR}^{(\rm I)}+
\underbrace{\M_{\rm SR}^{(\rm II)}+\M_{\rm IC}}_{\boldsymbol\Delta\M}.
\eeq

There is a remarkable parallelism between the decomposition
(\ref{eq:sr_one_two})
of the Wannier self-rotation (\ref{eq:Msr}) and the decomposition\cite{mv97}
\beq
\label{eq:om_partition}
\Omega=\Omega_{\rm I}+\widetilde\Omega
\eeq
of the Wannier spread
\beq
\label{eq:om}
\Omega=\sum_i^{\rm occ}\,\bra{w_i}(\hat\r-\overline\r_i)^2\ket{w_i}
\eeq
into a gauge-invariant part
\beq
\label{eq:om_i}
\Omega_{\rm I}=\sum_{\alpha}\,{\rm Tr}\,[\P\hat r_\alpha\Q \hat r_\alpha]
\eeq
and a gauge-dependent part
\beq
\label{eq:om_tilde}
\widetilde\Omega=\sum_{i,j\not= i}^{\rm occ}\,|\r_{ij}|^2 .
\eeq
The similarities between Eqs.~(\ref{eq:srone}) and (\ref{eq:om_i}), and
between Eqs.~(\ref{eq:srtwo}) and (\ref{eq:om_tilde}), are striking.
Interestingly, the gauge-invariant spread
$\Omega_{\rm I}$ is related to the ``ordinary'' absorption spectrum by a
second sum rule, as discussed in Ref.~\onlinecite{souza00} and
Appendix~\ref{app:other}.
In addition, the interpretation of $\Omega_{\rm I}$ as a measure of the
quadratic quantum fluctuations, or ``quantum spread,''
of the many-electron center of mass\cite{souza00} is mirrored
by $\srone$ having the meaning of a center-of-mass circulation, as
discussed in Appendix~\ref{app:many_body}.

First-principles calculations show that for maximally-localized WFs,
$\widetilde\Omega$ is typically much smaller than $\Omega_{\rm I}$.\cite{mv97}
Indeed, the minimization of the spread acts precisely to reduce
$\widetilde\Omega$ as much as possible.  In general
$\widetilde\Omega$ cannot be made to vanish exactly in two or higher
dimensions, since the non-commutativity of $\P\hat x\P$,  $\P\hat y\P$,
and $\P\hat z\P$ implies that the off-diagonal $\r_{ij}$ cannot all be zero.
In practice, however, they can become quite small.
According to Eq.~(\ref{eq:srtwo}), $\srtwo$ would also vanish if all
off-diagonal $\r_{ij}$ were precisely zero. Hence
we expect the self-rotation of maximally-localized WFs to be
dominated by the gauge-invariant part as well.\cite{explan-offdiag}

The fact that $\deltaM$ is composed of self-rotation and
itinerant-circulation parts which are not separately gauge-invariant
means that angular momentum can be converted back and forth between
$\sr$ and $\ic$ via gauge transformations.
This will be discussed in more detail in Sec.~\ref{sec:gauge};
here we simply note that the two parts are similar in that both originate
from the spatial overlap between neighboring
WFs. This is evident from the definition of $\srtwo$, and for
$\ic$ it follows from writing $\overline\v_i$ in terms of the ``current donated
from one Wannier orbital to its neighbors'' as in Ref.~\onlinecite{timo05}.
$\deltaM$ can therefore be interpreted as an {\it interorbital}
contribution to $\Morb$, even though it includes part of
the self-rotation, while $\srone$ is the purely {\it intraorbital} portion.
(Similarly, $\Omega_{\rm I}$ and $\widetilde\Omega$ are the intraorbital and
interorbital parts of the Wannier spread, respectively.)

\section{Crystalline solids}
\label{sec:bulk}

In this Section we apply the general formalism of Sec.~\ref{sec:bounded} to
crystalline solids, recasting the relevant quantities in the form of Brillouin
zone integrals. We start in
Sec.~\ref{sec:mcd_sum_rule_bulk} by rederiving the dichroic $f$-sum rule
for Bloch electrons.
In the remaining subsections we explore the connections between
this bulk reformulation and the theory of orbital magnetization in
crystals.\cite{timo05,xiao05,ceresoli06,shi07}

A somewhat unsatisfying aspect of that theory as developed in
Ref.~\onlinecite{ceresoli06} is the lack of consistency in the way
the orbital magnetization was decomposed, in the following sense.
One partition
($\Morb={\bf M}_{\rm LC}+\ic$ in their notation\cite{explan-notation})
was made for bounded samples, after which the thermodynamic
limit was taken for each term separately. The resulting $k$-space
expressions were then combined to form the total $\Morb$. Finally, working
in $k$-space, a {\it different} partition
($\Morb=\widetilde{\bf M}_{\rm LC}+\widetilde{\bf M}_{\rm IC}$)
was identified whose
individual terms were gauge-invariant, unlike those of the original
decomposition. In the process, however, the intuitive real-space
interpretation of the original decomposition was lost, and the
separate meanings of the two terms in the gauge-invariant
decomposition was left unclear.

Here, instead, we shall work from the very beginning with the two
gauge-invariant terms $\srone$ and $\deltaM$, which afford a simple
real-space interpretation in terms of WFs. They are first identified for
fragments with a crystalline interior (crystallites) in
Sec.~\ref{sec:relation_morb_bulk}. The thermodynamic limit of each term
is then taken, producing the reciprocal-space expressions of
Eqs.~(\ref{eq:sr1_bloch_gi})--(\ref{eq:deltaM_bloch_gi})
(the details of the derivation can be found in
Appendix~\ref{app:thm_limit}). 
Interestingly, we find
that our gauge-invariant terms
$\srone$ and $\deltaM$ differ from --~but are simply related to~-- those
of the gauge-invariant decomposition of Ref.~\onlinecite{ceresoli06}.
In the particular case of an insulator with a single valence band,
on the other hand, they reduce exactly to the terms identified in
Ref.~\onlinecite{xiao05}, as will be discussed in Sec.~\ref{sec:one_band}.
Because the work of Ref.~\onlinecite{xiao05}
is based on a semiclassical picture of
wavepacket dynamics, however, it is not easily generalized to a
multiband gauge-invariant framework as is done here.

In Eq.~(\ref{eq:m_orb_decomp}) of Sec.~\ref{sec:bounded} the
decomposition $\deltaM=\srtwo+\ic$ for insulating systems was obtained by
working in the Wannier representation.  For insulating crystallites
$\ic$  can be divided further into a ``surface'' part $\icsurf$
and an ``interior'' part $\icint$.
The interplay between the resulting three contributions to $\deltaM$ will be
the focus of the final two subsections.
Single-band insulators are discussed in Sec.~\ref{sec:one_band}.
The general case of multiband insulators is considered in
Sec.~\ref{sec:gauge}, where the gauge-transformation properties of those
terms is analyzed.

\subsection{Dichroic $f$-sum rule}
\label{sec:mcd_sum_rule_bulk}

The first step is to rewrite the Kubo-Greenwood formula (\ref{eq:sigma}) in a 
form
appropriate for periodic crystals, where dipole transitions connect
valence and conduction Bloch states with the same crystal momentum $\k$.
Eq.~(\ref{eq:f}) becomes, dropping the index $\k$ for conciseness,
\beq
\label{eq:f_bulk}
f_{nm,\alpha\beta}=
-(2m_e\omega_{mn}/\hbar)\langle u_n|\partial_\alpha u_m\rangle
\langle u_m|\partial_\beta u_n\rangle,
\eeq
where $\partial_\alpha\equiv \partial/\partial k_\alpha$ and we have used the 
relation\cite{ksv93} 
$v_{nm,\alpha}=\omega_{mn}\langle u_n|\partial_\alpha u_m\rangle$ for 
$m\not=n$, with $\ket{u_n}$ a cell-periodic Bloch state. Eq.~(\ref{eq:sigma}) 
now reads
\begin{equation}
\label{eq:sigma_bulk}
\sigma_{\rm H}(\omega)=\frac{\pi e^2}{2m_e}
\int\frac{d\k}{(2\pi)^3}\sum_n^{\rm occ}\sum_m^{\rm empty}
f_{nm}\delta(\omega-\omega_{mn}).
\end{equation}

Consider the frequency integral of $\sigma_{\rm H}(\omega)$,
\beq
\label{eq:mcd_sum_a}
\langle\sigma_{\rm H}\rangle=\frac{\pi e^2}{2m_e}
\int\frac{d\k}{(2\pi)^3}\,\sum_n^{\rm occ}\sum_m^{\rm empty}f_{nm}.
\eeq
The dichroic $f$-sum rule will be obtained from the imaginary part of this
complex quantity, while the real part yields the ordinary $f$-sum rule (see
Appendix~\ref{app:other}).

Using Eq.~(\ref{eq:f_bulk}) to expand the summation,
\begin{eqnarray}
\label{eq:f_summation}
&&\sum_n^{\rm occ}\sum_m^{\rm empty}f_{nm,\alpha\beta}=\nn
&=&-\frac{2m_e}{\hbar^2}\sum_n^{\rm occ}\sum_m^{\rm empty}\,
\langle u_n|\partial_\alpha u_m\rangle(E_m-E_n)
\langle u_m|\partial_\beta u_n\rangle\nn
&=&-\frac{2m_e}{\hbar^2}\sum_n^{\rm occ}\sum_m^{\rm empty}\,
\langle \partial_\alpha u_n| u_m\rangle(E_n-E_m)
\langle u_m|\partial_\beta u_n\rangle\nn
&=&\frac{2m_e}{\hbar^2}(g_{\k,\alpha\beta}-h_{\k,\alpha\beta}),
\end{eqnarray}
where we have introduced a set of notations as follows:
\begin{equation}
\label{eq:b}
b_{\k,\alpha\beta}=\sum_n^{\rm occ}\,\langle \ppartial_\alpha u_n|\ppartial_\beta u_n
\rangle,
\end{equation}
\begin{equation}
\label{eq:g}
g_{\k,\alpha\beta}=\sum_n^{\rm occ}\,\bra{\ppartial_\alpha u_n}\hat H
\ket{\ppartial_\beta u_n},
\end{equation}
and
\begin{equation}
\label{eq:h}
h_{\k,\alpha\beta}= \sum_n^{\rm occ}\,E_n \,
\langle \ppartial_\alpha u_n|\ppartial_\beta u_n \rangle.
\end{equation}
The symbol $\ppartial$ denotes the covariant
derivative,\cite{souza04,ceresoli06} defined as
$\ket{\ppartial_\alpha u_{n\k}}=\Q_\k\ket{\partial_\alpha u_{n\k}}$, where
$\hat Q_\k=\sum_m^{\rm empty}|u_{m\k}\rangle\langle u_{m\k}|$. The imaginary
part of $b_{\k,\alpha\beta}$ is essentially the Berry curvature while its real
part is related to the quantum metric 
(Appendix~C of Ref.~\onlinecite{mv97};
we discuss the physical content of $b_{\k,\alpha\beta}$ in
Appendix~\ref{app:other}).  Quantities $g_{\k,\alpha\beta}$
and $h_{\k,\alpha\beta}$ are similar to $b_{\k,\alpha\beta}$
except that they carry an extra factor of Hamiltonian or energy.
Note that $b_{\k,\alpha\beta}$ corresponds to the quantity
$f_{\k,\alpha\beta}$ in Ref.~\onlinecite{ceresoli06},
while  $g_{\k,\alpha\beta}$ and $h_{\k,\alpha\beta}$ are the same
as in that work.

With these definitions Eq.~(\ref{eq:mcd_sum_a}) becomes
\begin{equation}
\label{eq:sum_rule_hermitian}
\langle\sigma_{{\rm H},\alpha\beta}\rangle=
\frac{\pi e^2}{\hbar^2}
\int\frac{d\k}{(2\pi)^3}\,(g_{\k,\alpha\beta}-h_{\k,\alpha\beta}).
\end{equation}
The imaginary part reads, in vector form,
\begin{equation}
\label{eq:om_sum_rule_bloch}
\langle\boldsymbol{\sigma}''_{\rm A}\rangle=
\frac{\pi e^2}{\hbar^2}
\int\frac{d\k}{(2\pi)^3}\,
({\bf g}''_\k-{\bf h}''_\k).
\end{equation}
This is the dichroic $f$-sum rule in the Bloch representation.

We can now compare this result with the decomposition obtained
in Ref.~\onlinecite{ceresoli06}, where the
ground-state orbital magnetization was partitioned into two
gauge-invariant terms as
\begin{equation}
\label{eq:m_orb_ceresoli}
{\bf M}_{\rm orb}=\widetilde{\bf M}_{\rm LC}+\widetilde{\bf M}_{\rm IC}
\end{equation}
where
\begin{equation}
\label{eq:m_lc}
\widetilde{\bf M}_{\rm LC}=\frac{e}{\hbar c}\int\frac{d\k}{(2\pi)^3}\,
{\bf g}''_\k,
\end{equation}
\begin{equation}
\label{eq:m_ic}
\widetilde{\bf M}_{\rm IC}=\frac{e}{\hbar c}\int\frac{d\k}{(2\pi)^3}\,
{\bf h}''_\k.
\end{equation}
We thus arrive at our main result
\begin{equation}
\label{eq:om_sum_rule}
\langle\boldsymbol{\sigma}''_{\rm A}\rangle=\frac{\pi ec}{\hbar}
\big(\widetilde{\bf M}_{\rm LC}-\widetilde{\bf M}_{\rm IC}\big)
\end{equation}
relating the integrated MCD spectrum to the components of the orbital
magnetization.  Note that the sum rule is proportional to the
{\it difference} between the gauge-invariant contributions of
Ref.~\onlinecite{ceresoli06}. By independently measuring the sum
of $\widetilde{\bf M}_{\rm LC}$ and $\widetilde{\bf M}_{\rm IC}$
via gyromagnetic experiments\cite{kittel53} and the difference
via the magneto-optical sum rule, the value of each individual term
can indeed be measured in principle, resolving an open problem
posed in Ref.~\onlinecite{ceresoli06}.

Strictly speaking, Eqs.~(\ref{eq:m_orb_ceresoli})--(\ref{eq:m_ic}) as written
are valid for conventional insulators only. The generalization to metals and
Chern insulators is subtle, but the understanding emerging from
Refs.~\onlinecite{xiao05,ceresoli06,shi07} is that the appropriate
generalization is obtained by making the replacements $H\rightarrow
H-\mu$ and $E_n\rightarrow E_n-\mu$ in Eqs.~(\ref{eq:g}) and
(\ref{eq:h}), where $\mu$ is the electron chemical potential.  Clearly
$g_\k-h_\k$, and with it the sum rule (\ref{eq:om_sum_rule_bloch}),
are insensitive to these substitutions.\cite{explan-intra}

Comparing Eqs.~(\ref{eq:m_orb_ceresoli}) and
(\ref{eq:om_sum_rule}) for extended crystals
with Eqs.~(\ref{eq:M_decomp})  and (\ref{eq:mcd_sum_rule_bounded_b})  for
bounded samples, it appears plausible that the two partitions
(\ref{eq:M_decomp}) and (\ref{eq:m_orb_ceresoli}) of $\Morb$ ought to
be related by
\beq
\label{eq:sr1_bloch}
\srone=\widetilde\M_{\rm LC}-\widetilde\M_{\rm IC},
\eeq
\beq
\label{eq:deltaM_bloch}
\deltaM=2\widetilde\M_{\rm IC},
\eeq
or explicitly,
\beq
\label{eq:sr1_bloch_gi}
\srone=\frac{e}{\hbar c}\int\frac{d\k}{(2\pi)^3}\,({\bf g}''_\k-{\bf h}''_\k),
\eeq
\beq
\label{eq:deltaM_bloch_gi}
\deltaM=\frac{2e}{\hbar c}\int\frac{d\k}{(2\pi)^3}\,{\bf h}''_\k.
\eeq
The correctness of these identities is demonstrated in
Appendix~\ref{app:thm_limit} by taking the thermodynamic limit
of results derived in the next subsection.

\subsection{The magnetization of an insulating crystallite}
\label{sec:relation_morb_bulk}

To gain a better understanding of the bulk expressions derived 
in the previous section, we
now specialize the results obtained for bounded samples in
Sec.~\ref{sec:relation_morb} to the case that the sample has a
crystalline interior.  Working in the Wannier representation, we
are then able to establish connections between the $k$-space
and Wannier viewpoints and associate a local physical picture with
the various terms appearing in the bulk orbital magnetization.

Following Refs.~\onlinecite{timo05} and \onlinecite{ceresoli06}, we
divide our crystallite into ``surface'' and ``interior'' regions.
This division is largely arbitrary, and it only needs to satisfy two
requirements: (i) the border between the two regions
should be placed sufficiently deep inside the sample where
the local environment is already crystalline, and (ii) the surface region
should occupy a non-extensive volume in the thermodynamic limit.
The Wannier orbitals spanning the ground state are
assigned to each region. Those in the interior converge exponentially
to the bulk WFs $\ket{\R n}$ ($\R$ is a lattice
vector), and those on the surface will be denoted by $\ket{w_s}$.

We first divide the orbital magnetization into self-rotation (SR) and
itinerant-circulation (IC) contributions according to
Eqs.~(\ref{eq:sr_ic})--(\ref{eq:Mic}).
In the thermodynamic limit the SR part, which only involves the relative
coordinate $\hat\r-\overline\r_i$, is dominated by the interior region.
Invoking translational invariance,
\beq
\label{eq:sr_wann}
\sr=\facc\sum_n\,
\left[
\bra{\0 n}\hat\r\times\hat\v\ket{\0 n}-\overline\r_n\times\overline\v_n
\right],
\eeq
where $\facc=-e/(2cV_c)$, with $V_c$ the cell volume,
$\overline\r_n=\bra{\0 n}\hat\r\ket{\0 n}$,
and $\overline\v_n=\bra{\0 n}\hat\v\ket{\0 n}$.
Henceforth summations over
band-like indices span the valence-band states.

Next we break down the self-rotation as in Eq.~(\ref{eq:sr_one_two}), setting
$\P=\sum_\R\sum_n\,\ket{\R n}\bra{\R n}$:
\beq
\label{eq:sr1_wann}
\srone=\facc{\rm Re}\,{\rm tr}_c[\P\hat\r\times\Q\hat\v]
\eeq
%
%
\bea
\label{eq:sr2_wann}
\srtwo&=&\facc
\Big(
  {\rm Re}\,{\rm tr}_c[\P\hat\r\times\P\hat\v]-
  \sum_n\,\overline\r_n\times\overline\v_n
\Big)\nn
&=&\facc\sum_n\sum_{\R m\not=\0 n}
{\rm Re}\,\big\{
             \bra{\0 n}\hat\r\ket{\R m}\times\bra{\R m}\hat\v\ket{\0 n}
          \big\}.\nn
\eea
The symbol ${\rm tr}_c$ denotes the trace per unit cell.
Note that we have taken the real part of the traces explicitly;
this was not needed in Eqs.~(\ref{eq:srone}) and (\ref{eq:srtwo}) for
bounded samples, where the traces were automatically real.

Now we turn to the IC term (\ref{eq:Mic}) in Eq.~(\ref{eq:sr_ic}).
Unlike $\sr$, in the thermodynamic limit it generally has contributions from
both interior {\it and} surface regions:\cite{timo05,ceresoli06}
\beq
\label{eq:ic_wann}
\M_{\rm IC}=\M_{\rm IC}^{(\rm int)}+\M_{\rm IC}^{(\rm surf)}.
\eeq
The interior part becomes
\beq
\label{eq:ic_int_wann}
\M_{\rm IC}^{(\rm int)}
=\facc\sum_n\,\overline\r_n\times\overline\v_n
\eeq
where it was necessary to use
\beq
\label{eq:zero_current}
\sum_n\,\overline\v_n=0
\eeq
when exploiting the translational invariance.  Eq.~(\ref{eq:zero_current})
expresses the fact that
no macroscopic current, or dynamic polarization,\cite{souza04}  flows through
the bulk in a stationary state. Because of this constraint, the quantity
(\ref{eq:ic_int_wann}) necessarily vanishes for insulators with a single
valence band. In
multiband insulators it takes the form of an {\it intracell}
itinerant circulation: the WF centers in each
cell can have a net circulation while their collective center-of-mass remains
at rest.

Finally, the surface contribution is
\beq
\label{eq:ic_surf_wann}
\icsurf=\fac\sum_{s=1}^{N_s}\,\overline\r_s\times\overline\v_s.
\eeq
It was shown in Refs.~\onlinecite{timo05} and \onlinecite{ceresoli06}
that in the thermodynamic limit this can be recast as
\beq
\label{eq:ic_surf_wann_b}
\icsurf=-\faccbar\,{\rm Im}\sum_{m n\R}\,\R\times\bra{m\0}\hat\r\ket{n\R}
\bra{n\R}\H\ket{m\0},
\eeq
where $\faccbar=\facc/\hbar$.
This result is remarkable in that it expresses a circulation in the surface
region solely in terms of matrix elements between the interior WFs, in a way
that does not depend on the precise location of the
boundary between the two regions (provided that the boundary
satisfies the two criteria mentioned earlier). We emphasize that it holds
for {\it crystalline} insulators only.

Whereas $\icint$ is an intracell-like term, in the bulk form
(\ref{eq:ic_surf_wann_b}) $\icsurf$ is seen to have an {\it intercell}
character, vanishing in the ``Clausius-Mossotti''
limit of zero overlap between WFs
belonging to different cells. The assignment of the bulk WFs to
specific cells is however not unique, and by making a different choice it is
possible to convert between ``intracell'' $\icint$ and ``intercell''
$\icsurf$. For this and other reasons                      to be detailed in
Sec.~\ref{sec:gauge}, the interior and surface parts of $\Morb$ are in general
not physically well-defined, even in crystalline insulators. Collecting terms, 
the full orbital magnetization reads
\beq
\label{eq:morb_frag}
\M_{\rm orb}=\M_{\rm SR}^{(\rm I)}+
\underbrace{\M_{\rm SR}^{(\rm II)}+
\M_{\rm IC}^{(\rm int)}+\M_{\rm IC}^{(\rm surf)}}_{\deltaM},
\eeq
which is similar to Eq.~(\ref{eq:m_orb_decomp}) except that the IC term
has been separated into interior and surface parts.

This Wannier-based decomposition of the magnetization of a
crystallite  follows closely that of
Ref.~\onlinecite{ceresoli06}. Two differences are worth noting.  First,
we have emphasized the distinction between Wannier self-rotation and itinerant
circulation. In Ref.~\onlinecite{ceresoli06} the emphasis was more on the
separation between the surface contribution $\icsurf$ 
(denoted by $\M_{\rm IC}$ in that work) and the interior
contribution $\M_{\rm LC}=\facc{\rm tr}_c[\P\hat\r\times\hat\v]$ containing the
net magnetic dipole density of the WFs in a crystalline cell. This ``local
circulation'' includes all of the self-rotation as well as the intracell part
of the itinerant circulation. In the present notation the decomposition of
Ref.~\onlinecite{ceresoli06} reads
\beq
\label{eq:morb_frag_ceresoli}
\M_{\rm orb}=\underbrace{\M_{\rm SR}^{(\rm I)}+
\M_{\rm SR}^{(\rm II)}+
\M_{\rm IC}^{(\rm int)}}_{\M_{\rm LC}}+\M_{\rm IC}^{(\rm surf)}.
\eeq
Note that for one-band insulators $\icint=0$, in which case the interior 
contribution coincides with the self-rotation, and the surface part with the 
itinerant circulation.\cite{timo05}
Second, by identifying a gauge-invariant part of the self-rotation, we have
been able to organize the four resulting terms into the two gauge-invariant
groups indicated in Eq.~(\ref{eq:morb_frag}).

The present viewpoint appears to be more useful for arriving at a simple 
physical picture for the sum rule. It has the additional advantage of being 
applicable to disordered and microscopic systems, for which the distinction 
between interior and surface  contributions loses meaning.  

\subsection{One-band insulators}
\label{sec:one_band}

We begin our discussion of $\Morb$ in insulators with a
single valence  band by considering the remainder $\deltaM$.
We saw in Sec.~\ref{sec:relation_morb_bulk} that, of the three terms
into which it is naturally decomposed in the Wannier representation, one of
them vanishes if there is only one WF per cell,
\beq
\label{eq:sr_oneband}
\icint=0.
\eeq
Remarkably, the two surviving terms become identical,
\beq
\label{eq:deltaM_oneband}
\srtwo=\icsurf=\frac{\deltaM}{2},
\eeq
and thus {\it individually} gauge-invariant. This follows from
Eqs.~(\ref{eq:sr1_bloch})--(\ref{eq:deltaM_bloch}) in the one-band limit.
Indeed, the quantities $\widetilde\M_{\rm LC}$ and $\widetilde\M_{\rm IC}$
therein were defined in
Ref.~\onlinecite{ceresoli06} in such a way that for one-band insulators
they reduce to the quantities $\M_{\rm LC}=\sr$ and $\icsurf$
in Eq.~(\ref{eq:morb_frag_ceresoli}). It is then
seen that Eqs.~(\ref{eq:sr1_bloch}) and (\ref{eq:deltaM_bloch}) correspond to
the first and second equalities in Eq.~(\ref{eq:deltaM_oneband}) respectively.
We emphasize that Eqs.~(\ref{eq:sr_oneband})--(\ref{eq:deltaM_oneband}) only
hold for crystalline WFs which respect the full translational symmetry of
the crystal. If, for instance, a larger unit cell is used (effectively
folding the Brillouin zone and turning the system into a multiband
insulator), the additional gauge freedom can be used to construct WFs
for which Eqs.~(\ref{eq:sr_oneband})--(\ref{eq:deltaM_oneband}) no longer hold.

Consider now the full orbital magnetization. For one-band insulators
the reciprocal-space expressions
(\ref{eq:sr1_bloch_gi})--(\ref{eq:deltaM_bloch_gi}) reduce to
\beq
\label{eq:sr1_bloch_gi_oneband}
\srone=\frac{e}{2\hbar c}\,{\rm Im}\,\int\frac{d\k}{(2\pi)^3}\,
\bra{\partial_\k u_\k}\times(\H_\k-E_\k)\ket{\partial_\k u_\k}
\eeq
and
\beq
\label{eq:deltaM_bloch_gi_oneband}
\deltaM=\frac{e}{\hbar c}\,\int\frac{d\k}{(2\pi)^3}\,
E_\k\,{\rm Im}\,\bra{\partial_\k u_\k}\times\ket{\partial_\k u_\k}.
\eeq
Their sum $\Morb$ is given by the right-hand-side of
Eq.~(\ref{eq:sr1_bloch_gi_oneband}) with $-E_\k$ replaced by $+E_\k$, which is
the expression originally obtained in
Refs.~\onlinecite{timo05} and \onlinecite{xiao05}. Moreover,
the individual contributions $\srone$ and $\deltaM$ coincide with those
identified in Ref.~\onlinecite{xiao05}. Instead, the derivation of
Ref.~\onlinecite{timo05} leads to the alternative --~but, for one-band
insulators, also gauge-invariant~--
partition into the ``interior'' and ``surface'' parts
$\M_{\rm LC}=\sr=\srone+\deltaM/2$ and $\icsurf=\deltaM/2$.

While the individual terms $\srone$ and $\deltaM$ agree, for single-band
insulators, with those of Ref.~\onlinecite{xiao05}, we interpret them
somewhat differently here.  Eq.~(\ref{eq:sr1_bloch_gi_oneband}) of
Ref.~\onlinecite{xiao05} had the meaning of an intrinsic magnetic
moment associated with the self-rotation of the carrier wavepackets.
According to the present derivation, that term is only {\it part} of the Wannier
self-rotation. As for Eq.~(\ref{eq:deltaM_bloch_gi_oneband}), in the
derivation of Ref.~\onlinecite{xiao05} it was seen to arise from a
Berry-phase correction to the electronic density of states, and was
subsequently claimed to be associated with a boundary current
circulation.\cite{xiao06} Instead, according to the present viewpoint only
{\it half} of it originates in the itinerant circulation $\icsurf$ of the
surface WFs, while the other half is ascribed to the remaining self-rotation
$\srtwo$ of the WFs in the bulk.

\subsection{Gauge transformations for multiband insulators}
\label{sec:gauge}

In multiband insulators all three terms $\srtwo$, $\icint$, and
$\icsurf$ can be nonzero. However, their individual values are not
physically meaningful, since a gauge transformation can redistribute the total 
$\deltaM$ among them.  In particular, it is interesting to consider gauge 
transformations that shift the location of a WF by a lattice vector.

A general gauge transformation takes the form \cite{mv97}
\beq
\label{eq:gauge}
\ket{u_{n\k}}\rightarrow\sum_m\,\ket{u_{m\k}}U_{mn\k}
\eeq
where $U_{\k}$ is an $N_b\times N_b$ unitary matrix in the band indices.
We assume that a transformation of this kind has already been applied
to transform from the Hamiltonian eigenstates at each $\k$ to a
set of states that are smooth in $\k$ from which the WFs are to be
constructed.  We can then interpose an additional diagonal gauge transformation
\beq
\label{eq:gauge_diag_disc}
\ket{u_{n\k}}\rightarrow e^{-i\k\cdot\R_n}\ket{u_{n\k}},
\eeq
where $\R_n$ is a real-space lattice vector; this has the effect of
shifting the location of WF $n$
by $\R_n$.  For a one-band insulator, or if $\R_n$ is the
same for all bands, this amounts to shifting the
choice of the ``home'' unit cell.  However, in the multiband case
different WFs can be shifted differently,
corresponding to the freedom in choosing which WFs ``belong'' to
the home unit cell.

\begin{figure}
\begin{center}
   \epsfig{file=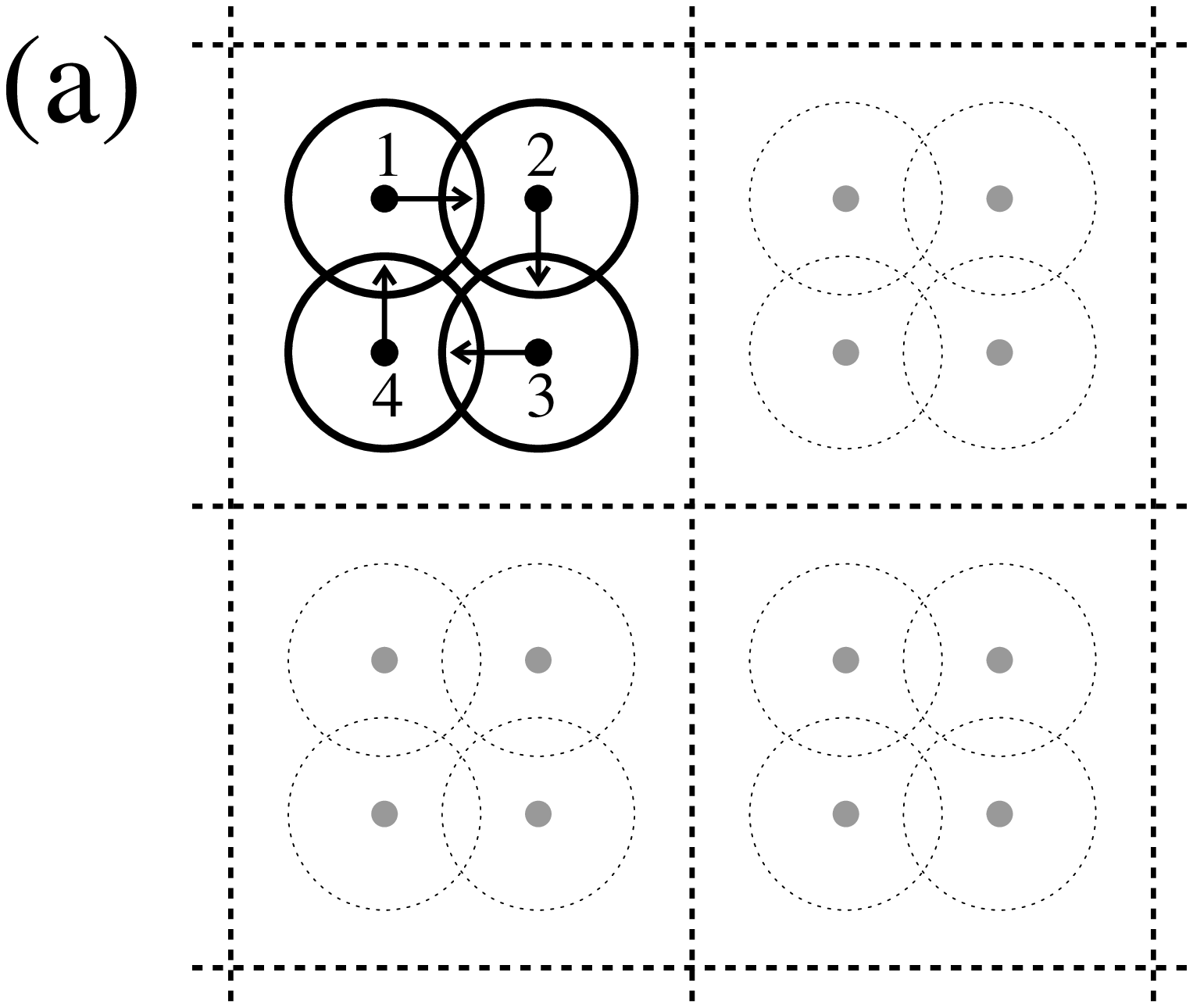,width=2.3in}
\vskip 0.5cm
   \epsfig{file=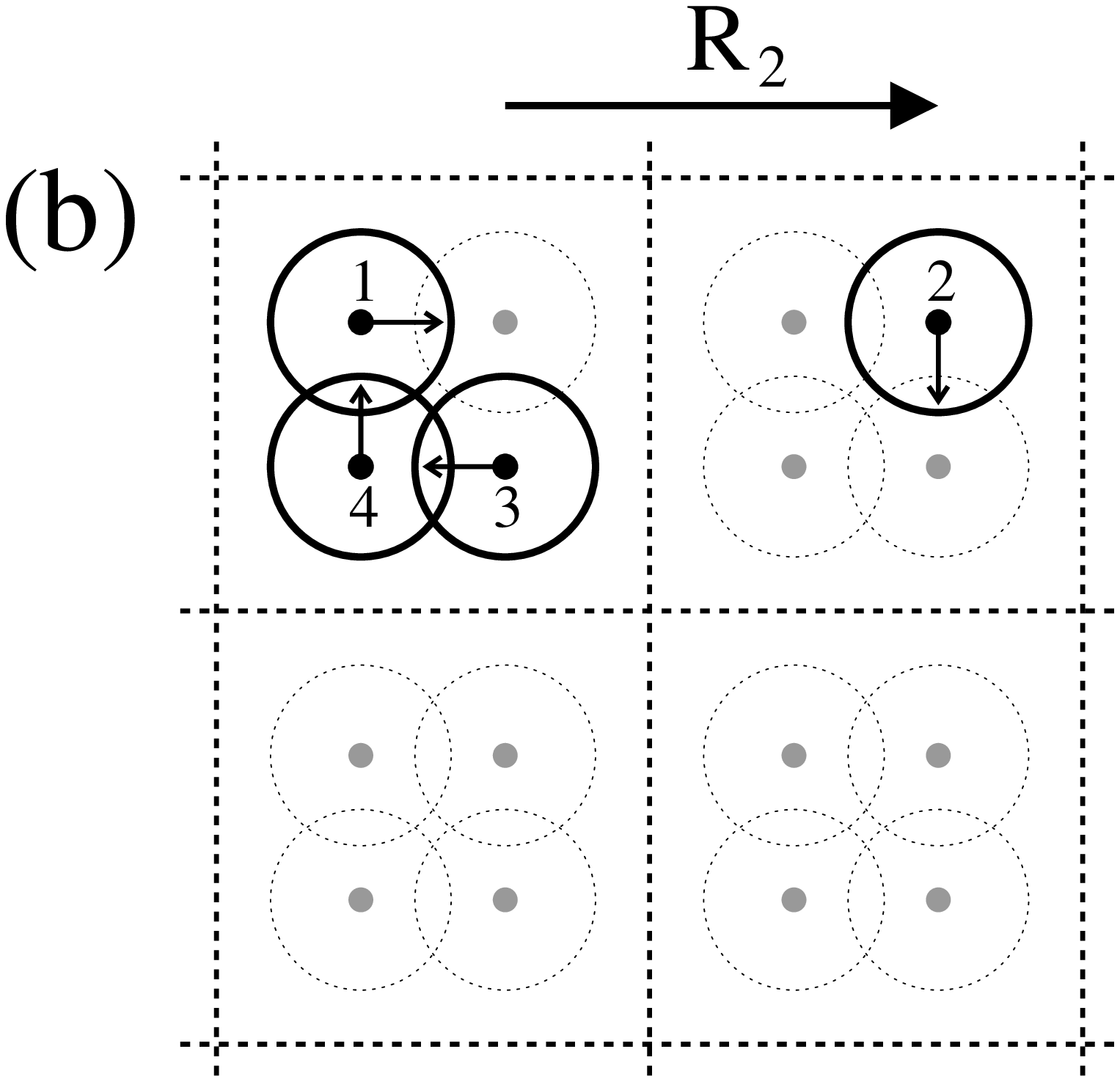,width=2.3in}
\end{center}
\caption{Schematic model of a molecular crystal with one
molecule per cell and four WFs per molecule.
The range of the orbitals is indicated by the overlapping circles, and their
center-of-mass velocities $\overline\v_n$ are denoted by arrows.
The two panels show in bold two possible choices of ``Wannier basis.''
}
\label{fig:IC}
\end{figure}

For example, Fig.~\ref{fig:IC} shows four cells of a model
two-dimensional crystal consisting of ``molecular magnets''
disposed on a square lattice with lattice constant $a$.
Before the transformation (\ref{eq:gauge_diag_disc}), the home unit
cell contains the four WFs shown in bold in Panel~(a).  Applying
the transformation with $\R_2=a\hat{x}$ and $\R_m=0$ for all other
WFs changes the selection of the ``basis'' of WFs
belonging to the home cell to be that shown in Panel~(b).

How does this affect the individual terms composing $\deltaM$?
Clearly the self-rotation (\ref{eq:sr_wann}) is not affected.
According to Eq.~(\ref{eq:ic_int_wann}), $\icint$  changes by
$\facc\R_2\times\overline\v_2$.
To preserve the overall invariance of $\deltaM$ the remaining term $\icsurf$
must change by an equal and opposite amount. Let us see in more detail
how this comes about.

We begin with a formal derivation. The $k$-space expression for
$\icsurf$ is given by\cite{ceresoli06}
\beq
\label{eq:icsurf}
\icsurf=\frac{e}{2\hbar c}{\rm Im}\,\sum_{mn}\,\int\frac{d\k}{(2\pi)^3}
E_{mn\k}\bra{\partial_\k u_{n\k}}\times\ket{\partial_\k u_{m\k}}.
\eeq
A few steps of algebra show that under the transformation
(\ref{eq:gauge_diag_disc}) $\icsurf$ changes by
\beq
\frac{e}{\hbar c}\R_n\times\sum_m\,\int\frac{d\k}{(2\pi)^3}{\rm Re}\,
\big\{
  \bra{u_n}\partial_\k u_m\rangle\bra{u_m}\hat H_\k \ket{u_n}
\big\}.
\eeq
Replacing $\bra{u_n}\partial_\k u_m\rangle$ by 
$-\bra{\partial_\k u_n} u_m\rangle$ allows to 
identify a term $\P_\k\hat{H}_\k=\hat{H}_\k$ in the above expression, which
becomes
\beq
\label{eq:gauge_icsurf}
-\frac{e}{\hbar c}\R_n\times\int\frac{d\k}{(2\pi)^3}
{\rm Re}
\big\{
  \bra{u_n}\H_\k\ket{\partial_\k u_n}
\big\}.
\eeq
Comparing with the Wannier velocity\cite{souza04}
\beq
\overline\v_n=-\frac{2V_c}{\hbar}\int\frac{d\k}{(2\pi)^3}
{\rm Re}
\big\{\bra{u_n}\H_\k\ket{\partial_\k u_n}\big\}
\eeq
and setting $n=2$ then produces the desired result 
$-\facc\R_2\times\overline\v_2$ for the change in $\icsurf$.

Coming back to the example in Fig.~\ref{fig:IC}, the intramolecular orbital
overlap gives rise to the nonzero velocities
$\overline \v_n$ indicated by the arrows.
With the choice of Wannier basis of Panel~(a), the collective circulation
of the Wannier centers in each cell results in a finite $\icint$, while from
Eq.~(\ref{eq:ic_surf_wann_b}) $\icsurf$ vanishes,
since there is negligible intercell overlap.
When the configuration of Panel~(b) is chosen, $\icsurf$
becomes $-\facc\R_2\times\overline\v_2$.  From
the present viewpoint this nonzero value is
made possible by the intramolecular (but now {\it inter}cell)
overlap between the second WF of each cell with WFs one, three, and four from
the cell shifted by $\R_2$.

\begin{figure}
\begin{center}
   \epsfig{file=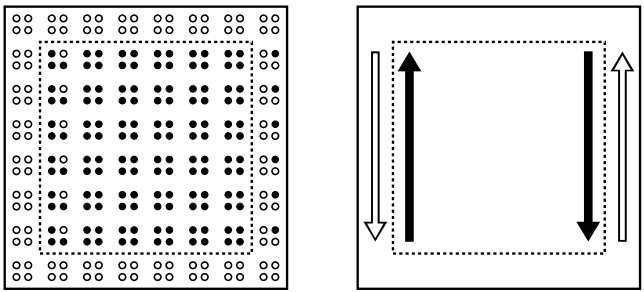,width=3.35in}
\end{center}
\caption{A finite sample cut from the bulk crystal of
Fig.~\ref{fig:IC}.
With the choice of Wannier basis of Fig.~\ref{fig:IC}(a),
``interior'' and ``surface'' WFs lie inside and outside the dashed
line respectively; with that of Fig.~\ref{fig:IC}(b),
they are denoted by solid and open circles respectively.
Right panel: open and solid arrows show the extra
``itinerant currents'' ($\icsurf$ and $\icint$ respectively)
associated with the latter choice.
}
\label{fig:IC_cluster}
\end{figure}

To view $\icsurf$ as a surface contribution rather than a bulk intercell term,
we consider now  a finite sample of the same crystal
(Fig.~\ref{fig:IC_cluster}),
which has been divided into surface and interior regions.
In deciding which WFs are ``interior-like'' and which are ``surface-like''
we shall require that all WFs assigned to the same cell must
belong to the same region. If the Wannier basis of
Fig.~\ref{fig:IC}(a) is used, the surface region can be chosen to
comprise the outermost layer of
molecules, so that the border between the two regions is given by
the dashed line.
The four WFs on each molecule form a unit with some internal
IC circulation but zero
center-of-mass velocity. The total sample magnetization is the sum of all such
internal circulations, which in the large-sample limit is interior-dominated,
so that $\icsurf\rightarrow 0$.

If the Wannier basis of Fig.~\ref{fig:IC}(b) is chosen instead, the upper
and lower surface regions are still composed of the outermost layer of
molecules. However, the left surface now contains, in addition, one
WF from each molecule in the second layer.
Those lone surface WFs carry a downward particle ``IC current'' which extends
along the left surface and is indicated by an open arrow on the right panel.
A corresponding IC current appears on the right surface, and
together they yield $\icsurf=-(e/2ca)|\overline\v_2|\hat{\bf z}$,
which agrees with the result $\icsurf=-\facc\R_2\times\overline\v_2$ found
earlier using a purely bulk argument [in this example $\facc=-e/(2ca^2)$].
A change of gauge should not change any physical quantity, such as the
actual current flowing on the left surface.  Since it appears to change by
adding the open arrow, there must be another equal and opposite
contribution (the adjacent solid arrow). This contribution is the ``interior''
IC current carried by the remaining three WFs (filled circles) on the
molecules of the second layer.

The situation just described is reminiscent of the
``quantum of polarization'' in the theory of
dielectric polarization,\cite{vks93}
where a change of Wannier basis like that
leading from Fig.~\ref{fig:IC}(a) to Fig.~\ref{fig:IC}(b) shifts
the polarization by a quantum and also changes the surface charge
by one electron per surface cell area.
This might suggest
that the full gauge invariance of the interior and surface parts of
$\Morb$ discussed in Sec.~\ref{sec:one_band} for
single-band insulators becomes, in multiband insulators, a
gauge-invariance modulo $\facc\R_n\times\overline\v_n$.
While true for this particular example, this is generally not
so.\cite{explan-quantum} Even for this model it will cease to be true
as soon as the molecules start overlapping
significantly. When this happens, the value of
$\icsurf$ can be tuned continuously using other types
of gauge transformations, e.g.,
the continuous diagonal transformation
\beq
\label{eq:gauge_diag_cont}
\ket{u_{n\k}}\rightarrow e^{i\theta_{n\k}}\ket{u_{n\k}}
\eeq
with $\theta_{n,\k+{\bf G}}=\theta_{n\k}$. 
This produces a change in $\icsurf$ given by
Eq.~(\ref{eq:gauge_icsurf}) with $\R_n$ therein replaced by a factor of
$-\partial_\k\theta_{n\k}$ in the integrand.
Since both $\overline\r_n$ and $\overline\v_n$ remain invariant
(the former was shown in Ref.~\onlinecite{ksv93} and the
latter follows from Eq.~(\ref{eq:zero_current})
together with the fact that all other
$\overline\v_m$ are unaffected), so does
$\icint$. The change in $\icsurf$ must therefore be absorbed by $\srtwo$.

To summarize, the transformation
(\ref{eq:gauge_diag_disc}) transfers discrete amounts of itinerant
circulation between the interior and surface regions, while the transformation
(\ref{eq:gauge_diag_cont})
converts continuously between interior self-rotation  and surface itinerant
circulation. Finally, under the most general transformation
(\ref{eq:gauge}) all three gauge-dependent terms in Eq.~(\ref{eq:morb_frag})
can be affected simultaneously, so that only their sum $\deltaM$ is unique and
physically meaningful.

\section{Summary and outlook}
\label{sec:summary}

We have presented an exact sum rule for the MCD spectrum,
elucidated its physical interpretation, and related it to the recent rigorous
formulation of orbital magnetization in crystals.
In insulating systems the sum rule
probes the gauge-invariant part $\srone$ of the self-rotation of the
occupied Wannier orbitals.
The total orbital magnetization has a second, less
obvious contribution
$\deltaM$, arising from the overlap between neighboring WFs. It
comprises both self-rotation (SR) and
itinerant-circulation (IC) parts in proportions which depend on the precise
choice of WFs, while $\deltaM$ itself has a unique value.
Although the intuitive interpretation in terms of the occupied WFs is
restricted to Wannier-representable systems such as
conventional insulators, the terms $\srone$ and $\deltaM$ are in fact
well-defined for all electron systems, including metals and Chern
insulators.

The practical importance of the sum rule is that it allows to break down
$\Morb$ into physically meaningful parts, using a combination
of gyromagnetic and magneto-optical measurements. This should provide
valuable information on the intraorbital (or localized) versus interorbital
(or itinerant) character of orbital magnetism. For example, it has
been suggested 
(Ref.~\onlinecite{yafet63},
Appendix~B) that the anomalously large $g$-factors of Bi
might be caused by itinerant circulations very much like
the ones discussed here. On the basis of the present work it should
now be possible to test this conjecture.

In the last decade and a half a
sum rule for the X-ray MCD spectrum\cite{thole92} has been used extensively to
obtain site-specific information about orbital magnetism in solids.
The resulting orbital moments have been compared with gyromagnetic
measurements  of $\Morb$.\cite{chen95} 
If a significant itinerant contribution $\deltaM$ is present,
one may expect a discrepancy between the XMCD orbital
moments and the total $\Morb$ inferred from gyromagnetics.
It would therefore be of great interest to find such
systems defying the conventional wisdom about the connection between
the MCD spectrum and orbital magnetization.

The ideas discussed in this work should be most relevant for materials
displaying appreciable orbital magnetism and, in particular, appreciable
{\it interorbital} effects which might enhance
the ratio $|\deltaM/\srone|$.
These criteria do not favor band ferromagnets.
First, their orbital magnetization tends to be relatively small.
In the transition metal ferromagnets Fe, Co, and Ni, for example, it accounts
for less than 10\% of the spontaneous magnetization.\cite{kittel53}
(For comparison, the field-induced orbital magnetization of the $d$
paramagnetic metals can be as large as the induced spin
magnetization. This has been established both from gyromagnetic
experiments\cite{huguenin71} and from first-principles 
calculations.\cite{hjelm95})
Secondly, ferromagnetism is favored by narrow bands and localized orbitals,
for which interorbital effects are expected to be relatively minor.
Finally, the spin-orbit-induced $\Morb$ of ferromagnets
is believed to be an essentially atomic phenomenom largely confined to
a small core region close to the nucleus,\cite{solovyev98,solovyev05} and one 
might therefore expect $\deltaM$ to be small.
Paramagnets and diamagnets, with relatively wide bands 
(e.g., the $s$-$p$ metals and semiconductors) 
and additional contributions to $\Morb$ unrelated to 
spin-orbit, therefore appear to be more promising candidates.
Among ferromagnets, the 
``zero magnetization ferromagnets,''\cite{adachi99} 
whose
orbital magnetization is so large as to cancel the spin magnetization, might
be particularly interesting.

An important direction for future work is to carry out
first-principles calculations of $\srone$ and $\deltaM$ for
real materials. 
Such calculations would test the validity of the assumption
that orbital magnetism in solids is atomic-like in 
nature.\cite{solovyev98,solovyev05}
While plausible,
that assumption was made in the past partly out of practical necessity, since a
rigorous bulk definition of $\Morb$ in terms of the extended Bloch states
was not available. 
Confronting experiment with a
full calculation of $\Morb$ within spin-density functional theory (SDFT),
including the itinerant terms, 
would clarify whether SDFT can adequately
describe orbital magnetism in solids, or whether
an extended framework (e.g., LSD+U including 
``orbital polarization'' terms\cite{solovyev98,solovyev05} 
or current- and spin-density
functional theory\cite{shi07}) is needed.

In Appendix~A we place the dichroic $f$-sum rule in the broader context
of other known sum rules. We note in particular that by taking different
inverse-frequency moments, the interband MCD spectrum can be related to two 
other phenomena resulting from broken time-reversal symmetry, namely
the ground state orbital magnetization and the intrinsic anomalous
Hall effect.
These are generally expected to coexist, and this is
indeed the case for ferromagnets, where all three occur spontaneously.
In the case of Pauli paramagnets, however,
the intrinsic Hall mechanism of Karplus and Luttinger has received little 
if any attention. On the other hand, it is known that Pauli paramagnets
can display 
a field-induced MCD spectrum.\cite{yaresko98,ebert03}
This raises the question as to what role  the Berry curvature
may play in their ``ordinary'' (field-induced)
Hall effect. Such a ``dissipationless'' contribution is undoubtedly present in 
principle by virtue of the sum rule (\ref{eq:ihc_sum_rule}).
First-principles calculations of this effect will be
presented in a future communication.\cite{yates-unpublished}

To conclude, we have described the orbital magnetization of crystals in 
terms 
of localized ($\srone$) and itinerant ($\deltaM$) parts, and shown how to
relate them to magneto-optical and gyromagnetic  observables.
This should allow one to probe more deeply into the nature of
magnetism in condensed-matter systems than previously possible.

\acknowledgments

This work was supported by NSF Grant DMR-0549198.

\appendix

\section{Other sum rules}
\label{app:other}

In this Appendix we derive three additional sum rules for Bloch electrons and
discuss their relation to the dichroic $f$-sum rule. All four involve
inverse-frequency moments $\langle\omega^{-p}\sigma_{\rm H}\rangle$ [in the
notation of Eq.~(\ref{eq:bracket})] of the absorption spectrum
(\ref{eq:sigma_h}). They are given by $p=0$ and $p=1$, and in each case two 
sum rules are obtained by taking the real and imaginary parts: one ordinary 
and the other dichroic, respectively.

We first consider $p=0$.  From
the imaginary part of Eq.~(\ref{eq:sum_rule_hermitian}) we obtained the
dichroic $f$-sum rule (\ref{eq:om_sum_rule}). To discuss the real part we
revert from (\ref{eq:sum_rule_hermitian}) to the form (\ref{eq:mcd_sum_a}),
\beq
\label{eq:f-sum_ord_form}
\langle \sigma'_{\rm S,\alpha\beta}\rangle=
\frac{\pi e^2}{2m_e}\int\frac{d\k}{(2\pi)^3}
\sum_n^{\rm occ}\sum_m^{\rm empty}\,f'_{nm,\alpha\beta}.
\eeq
Since $f_{nm,\alpha\beta}=-[f_{mn,\alpha\beta}]^*$,
\bea
\label{eq:manifold_additive}
\sum_n^{\rm occ}\sum_m^{\rm empty}\,f'_{nm,\alpha\beta}&=&
\sum_n^{\rm occ}\sum_{m\not=n}\,f'_{nm,\alpha\beta}\nn
&=&
\sum_n^{\rm occ}\left[\delta_{\alpha\beta}-
\left(\frac{m_e}{m_e^*}\right)_{n,\alpha\beta}\right],
\eea
where the second equality is the effective-mass theorem. Hence we find
\beq
\label{eq:f-sum_ord}
\langle \sigma'_{{\rm S},\alpha\beta}\rangle=
\frac{\pi e^2}{2m_e}\int\frac{d\k}{(2\pi)^3}\sum_n^{\rm occ}\,
\left[
  \delta_{\alpha\beta}-\left(\frac{m_e}{m_e^*}\right)_{n,\alpha\beta}
\right],
\eeq
the {\it modified $f$-sum rule}\cite{mott_jones} for the ordinary spectrum.

To obtain the two sum rules for $p=1$ we again
start from Eq.~(\ref{eq:sigma_bulk}), but now replace
Eq.~(\ref{eq:f_summation}) by
\beq
\label{eq:first_moment}
\sum_n^{\rm occ}\sum_m^{\rm empty}\frac{f_{nm,\alpha\beta}}{\omega_{mn}}=
\frac{2m_e}{\hbar}\,b_{\k,\alpha\beta},
\eeq
where $b_{n,\alpha\beta}$ was defined in Eq.~(\ref{eq:b}). Thus
\beq
\label{eq:one_sum_rule_form}
\langle\omega^{-1}\sigma_{\rm H}\rangle=
\frac{\pi e^2}{\hbar}
\int\frac{d\k}{(2\pi)^3}\,b_\k.
\eeq

For the dichroic part, noting
that $\boldsymbol{\Omega}_\k=-2{\bf b}''_\k$ is the Berry curvature summed
over the occupied states at $\k$, and comparing with the ``intrinsic''
Karplus-Luttinger Hall conductivity\cite{yao04}
\beq
\label{eq:ihc}
\boldsymbol{\sigma}'_{\rm A}(\omega=0)=-\frac{e^2}{\hbar}
\int\frac{d\k}{(2\pi)^3}\,\boldsymbol{\Omega}_\k,
\eeq
one finds the {\it Hall sum rule},
\beq
\label{eq:ihc_sum_rule}
\langle\omega^{-1}\boldsymbol{\sigma}''_{\rm A}\rangle=
\frac{\pi}{2}\boldsymbol{\sigma}'_{\rm A}(\omega=0).
\eeq
This is the $\omega\rightarrow 0$ limit of the
Kramers-Kronig relation for the antisymmetric
conductivity.\cite{bennett65}
Since only the {\it interband} part of the optical
conductivity was included on the left-hand-side, the
{\it intrinsic} dc Hall conductivity was obtained on the
right-hand-side. Extrinsic
contributions to the latter (e.g., skew scattering) 
presumably arise from intraband terms in the former.

Finally consider the ordinary (real) part of
Eq.~(\ref{eq:one_sum_rule_form}).
The quantity $b'_{n,\alpha\beta}$ is the quantum metric.\cite{mv97} It is related
to the localization tensor $\Lambda_{\alpha\beta}$ of insulators
by\cite{resta02}
\beq
\label{eq:lambda}
\Lambda_{\alpha\beta}=\frac{V}{N}\int\frac{d\k}{(2\pi)^3}\,
b'_{\k,\alpha\beta},
\eeq
where $N/V$ is the electron density. Hence we recover the
{\it electron localization sum rule}\cite{souza00}
\beq
\label{eq:loc_sum_rule}
\langle\omega^{-1}\sigma'_{\rm S,\alpha\beta}\rangle=
\frac{\pi e^2N}{\hbar V}\Lambda_{\alpha\beta}.
\eeq

In summary, we have in Eqs.~(\ref{eq:sum_rule_hermitian}) and
(\ref{eq:one_sum_rule_form}) two general sum rules for the
zero-th and first inverse frequency moments of the optical absorption,
respectively. Taking imaginary and real
parts of (\ref{eq:sum_rule_hermitian}) gives 
the dichroic $f$-sum
rule (\ref{eq:om_sum_rule}) and the modified ordinary $f$-sum rule
(\ref{eq:f-sum_ord}), while taking imaginary and real parts
of (\ref{eq:one_sum_rule_form}) gives 
the Hall
sum rule (\ref{eq:ihc_sum_rule}) and the electron localization sum
rule (\ref{eq:loc_sum_rule}).

Besides emerging from a unified formalism, the four sum rules display certain
similarities. For instance, it will be shown in Appendix~\ref{app:many_body}
that the dichroic $f$-sum rule yields the
expectation value of the many-electron center-of-mass circulation
operator, while the trace of the localization tensor yields the
spread of the center-of-mass quantum distribution.
Moreover, in a one-particle picture each quantity can be viewed as the
gauge-invariant part of the corresponding property (self-rotation or spread)
of the Wannier orbitals, as discussed in
Sec.~\ref{sec:relation_morb} for bounded systems.
There is however one important difference between
the behavior of the two quantities in the
thermodynamic limit. While the center-of-mass circulation remains well-defined
for metals, the trace of the localization tensor is only meaningful for
insulators, diverging in metals.\cite{souza00,resta02}
Interestingly, the delocalization of electrons in metals
is also responsible for a correction to the $f$-sum rule.
Contrary to the canonical $f$-sum rule for atoms,\cite{sakurai}
the modified $f$-sum rule (\ref{eq:f-sum_ord}) does not yield the number
density of valence electrons in a metal, due to the presence of the last
term on the right-hand-side. This term appears because the Bloch states are
extended and do not vanish at
infinity.\cite{mott_jones} The fact that the correction term nevertheless
vanishes for insulators
is a consequence of the localized nature of insulating many-body
wavefunctions in {\it configuration} space.\cite{kohn68}

We conclude by noting that Eq.~(\ref{eq:ihc_sum_rule})
provides an extreme example of how sum rules from atomic physics
can change qualitatively when applied to extended systems. Indeed, the
corresponding sum rule for bound systems produces a vanishing
result,\cite{explan-vanishing}
\beq
\label{eq:kuhn}
\langle\omega^{-1}\sigma''_{{\rm A},\alpha\beta}\rangle=
\frac{\pi e^2}{\hbar V}{\rm Im}\,{\rm Tr}[\P\hat r_\alpha\Q\hat r_\beta]=0.
\eeq
In contrast, the bulk formula (\ref{eq:ihc_sum_rule})
produces for Chern insulators a quantized Hall conductivity, and  it also
describes the intrinsic anomalous Hall conductivity of ferromagnetic
metals.\cite{yao04}
This apparent contradiction highlights the subtleties associated with
the process of taking the thermodynamic limit and switching from open
to periodic boundary conditions for non-Wannier-representable
systems. Such issues are still not fully resolved in the theory of orbital
magnetization. While a general derivation of the bulk formula for $\Morb$
has been given working from the outset with a
periodic crystal,\cite{shi07} derivations
which start from finite crystallites and take them to the thermodynamic  limit
(Refs.~\onlinecite{timo05} and \onlinecite{ceresoli06} and
Appendix~\ref{app:thm_limit}) are
presently restricted to conventional insulators.

\section{Dichroic $f$-sum rule and the many-body wavefunction}
\label{app:many_body}

In the main text we interpreted the dichroic
$f$-sum rule, and the associated decomposition (\ref{eq:M_decomp}) of
$\Morb$, in an independent-particle picture based on WFs.
It is also possible to relate
these quantities directly to properties of the many-electron wavefunction,
without invoking any particular single-particle representation.
In preparation for that, let us first discuss a
one-electron system (e.g., a hydrogen atom in a magnetic field).
Its absorption spectrum is composed of sharp lines,
and is more conveniently described in terms of an oscillator strength rather
than an optical conductivity. Taking the imaginary part of
Eq.~(\ref{eq:f}) and using the relation (\ref{eq:vel_pos})
to replace one of the velocity matrix elements,
\beq
f''_{nm,\alpha\beta}=-\frac{m_e}{\hbar}
\big[
  \bra{n}\hat r_\alpha\ket{m}\bra{m}\hat v_\beta\ket{n}-
   (\alpha\leftrightarrow\beta)
\big].
\eeq
Summing over $m\not=n$ and using the closure relation together with
$\bra{n}\hat\v\ket{n}=0$ one finds, in vector notation,
\beq
\label{eq:f_sum_micro}
\sum_{m\not= n}\,{\bf f}''_{nm}=-\frac{m_e}{\hbar}
\bra{n}\hat\r\times\hat\v\ket{n}.
\eeq
This is the original dichroic $f$-sum rule of Hasegawa and
Howard,\cite{hasegawa61} with the
orbital angular momentum appearing on the right-hand-side;
in the notation of Sec.~\ref{sec:mcd_sum_rule} it reads
$\langle\boldsymbol\sigma''_{\rm A}\rangle=(\pi ec/\hbar)
\,\M_{\rm orb}$ (since here $\deltaM=0$).

We now generalize the discussion to $N$-electron systems.  In this context
$\hat\r=\sum_{i=1}^N\hat\r_i$ and $\hat\v=\sum_{i=1}^N\hat\v_i$,
and it is crucial to make a distinction between the one-particle operator
$\hat{\mathbf{\Lambda}}^{(1)}=\sum_{i=1}^N\hat{\bf r}_i\times\hat{\bf v}_i$
and the two-particle operator $\hat{\mathbf{\Lambda}}^{(2)}=\hat\r\times\hat\v$
appearing in Eq.~(\ref{eq:f_sum_micro}), as emphasized in
Ref.~\onlinecite{kunes00}. The former is related to the electronic angular
momentum and orbital magnetization, while the latter is related to 
a many-electron center-of-mass circulation.
(In the classical context, for example, a pair of electrons orbiting
180$^\circ$ out of phase in the same circular orbit would have
${\mathbf{\Lambda}}^{(1)}\ne0$ but ${\mathbf{\Lambda}}^{(2)}=0$.)

The derivation of the dichroic sum rule for the $N$-electron case
proceeds as before, except that
the velocity matrix elements in Eq.~(\ref{eq:f}) become
$\v_{nm}=\bra{\Psi_n}\hat\v\ket{\Psi_m}$, where $\ket{\Psi_m}$ are
now many-body eigenstates.  The result is still given by
Eq.~(\ref{eq:f_sum_micro}), with $\ket{n}$ replaced by $\ket{\Psi_n}$.
Indeed, it is natural to define the many-body generalization of
Eq.~(\ref{eq:srone}) as
\beq
\label{eq:srone-mb}
\srone=\fac\bra{\Psi_n}\hat{\mathbf{\Lambda}}^{(2)}\ket{\Psi_m}
\eeq
so that Eq.~(\ref{eq:mcd_sum_rule_bounded_b}) continues to hold.  From
this many-body perspective the difference $\deltaM$ with respect to the
full $\Morb$ is seen to arise from the cross terms
$\sum_{i,j\not=i}^N\hat{\bf r}_i\times\hat{\bf v}_j$ in
$\hat{\mathbf{\Lambda}}^{(2)}-\hat{\mathbf{\Lambda}}^{(1)}$.

To recover from (\ref{eq:srone-mb})
the independent-particle expression (\ref{eq:srone}) we
specialize to the case where $\ket{\Psi_m}$ is a single Slater determinant.
In second-quantized notation
$\hat\r=\sum_{ij}\,\r_{ij}\c_ic_j$, $\hat\v=\sum_{ij}\,\v_{ij}\c_ic_j$,
and $\ket{\Psi_0}=\c_1\ldots\c_N\ket{0}$,
where $i$ and $j$ label orthogonal one-particle states.
Then Eq.~(\ref{eq:srone-mb}) becomes
\bea
\label{eq:srone_slater_a}
\M_{\rm SR}^{(\rm I)}&=&
\fac\big(\bra{0}c_N\ldots c_1\big)\Big(\sum_{ij}\,\r_{ij}\c_ic_j\Big)\times\\
&\times&\Big(\sum_{kl}\,\v_{kl}\c_kc_l\Big)
\big(\c_1\ldots\c_N\ket{0}\big).
\eea
Terms in which the indices do not pair can immediately be eliminated.
Furthermore, pairings of the form ($k=l$, $i=j$) give no contribution,
since this leads to
$(\sum_i^{\rm occ}\,\r_{ii})\times (\sum_k^{\rm occ}\,\v_{kk})$
which vanishes because $\langle\Psi_0|\hat\v|\Psi_0\rangle=0$.
The only surviving terms are those with ($j=k$, $i=l$), yielding
\bea
\label{eq:srone_slater_b}
\M_{\rm SR}^{(\rm I)}&=&\fac\sum_{ij}\,\r_{ij}\times\v_{ji}
\big(\bra{\Psi_0}c_jc_j^\dagger\big)
\big(c_i^\dagger c_i\ket{\Psi_0}\big)\nn
&=&\fac\sum_{ij}\,\r_{ij}\times\v_{ji}(1-n_j)n_i=
\fac\sum_i^{\rm occ}\sum_j^{\rm empty}\r_{ij}\times\v_{ji},\nn
\eea
where $n_i$ is the state occupancy. Clearly the expression on the
right-hand-side is equivalent to that in Eq.~(\ref{eq:M_decomp}).

A similar analysis can be made for the other sum rules presented in
Appendix~\ref{app:other}. For example, the counterpart of the Hall sum rule
for a bounded many-electron system reads
\beq
\label{eq:kuhn_many_el}
\sum_{m\not=0}\frac{{\bf f}''_{0m}}{\omega_{m0}}=
-\frac{im_e}{\hbar}\bra{\Psi_0}\hat\r\times\hat\r\ket{\Psi_0}=0,
\eeq
which was termed in Ref.~\onlinecite{smith76} the
{\it Kuhn sum rule}. The independent-particle form
(\ref{eq:kuhn}) can be recovered from (\ref{eq:kuhn_many_el}) along the lines 
of Eqs.~(\ref{eq:srone_slater_a})--(\ref{eq:srone_slater_b}).
As for the electron localization sum rule, it yields
the second cumulant-moment of the quantum distribution of the many-electron
center-of-mass.\cite{souza00} In the independent-particle limit this reduces
to $\Lambda_{\alpha\beta}=(1/N){\rm Tr}[\P \hat r_\alpha\Q \hat r_\beta]$, whose
trace yields the gauge-invariant WF spread (\ref{eq:om_i}).
The bulk formula (\ref{eq:lambda}) for insulating
crystals can be recovered in the thermodynamic limit following the
strategy described below for the orbital magnetization.

\section{Thermodynamic limit}
\label{app:thm_limit}

In this Appendix we start from the expressions (\ref{eq:srone}) and
(\ref{eq:deltaM}) for $\srone$ and $\deltaM$ of insulating
crystallites and, by taking the thermodynamic limit in the Wannier
representation, turn them into the reciprocal-space expressions
(\ref{eq:sr1_bloch_gi}) and (\ref{eq:deltaM_bloch_gi}).

Before proceeding, recall that the quantities ${\bf g}_\k$ and
${\bf h}_\k$ entering Eqs.~(\ref{eq:sr1_bloch_gi})--(\ref{eq:deltaM_bloch_gi})
were defined in Eqs.~(\ref{eq:g})--(\ref{eq:h}) in the context of the
``Hamiltonian gauge'' in which $n$ labels a Bloch energy eigenstate.
Here, we work with a generalized Wannier representation as in
Sec.~\ref{sec:relation_morb_bulk}, where $n$ labels a Wannier
function and $\ket{u_{n\k}}$ is the state of Bloch symmetry (generally not
an energy eigenstate) constructed from that Wannier function.\cite{mv97}
The two representations are related by a $k$-dependent
unitary rotation as in Eq.~(\ref{eq:gauge}).
Then Eq.~(\ref{eq:g}) remains valid in the present
context, since it already takes the form of a trace, while Eq.~(\ref{eq:h})
is now replaced by
\beq
\label{eq:h_gi}
h_{\k,\alpha\beta}=\sum_{nm}\,E_{nm\k}\bra{\ppartial_\alpha u_{m\k}}
\ppartial_\beta u_{n\k}\rangle,
\eeq
where $E_{nm\k}=\bra{u_{n\k}}\H_k\ket{u_{m\k}}$.
With ${\bf g}_\k$ and ${\bf h}_\k$ written as traces in this way, it
is evident that each is a gauge-invariant quantity.\cite{ceresoli06}

\subsection{Gauge-invariant self-rotation $\srone$}
\label{sec:gi_self}

For insulating crystallites in the thermodynamic limit,
Eq.~(\ref{eq:srone}) can be replaced by
Eq.~(\ref{eq:sr1_wann}).  Thus we need to establish the equivalence
between Eqs.~(\ref{eq:sr1_wann}) and (\ref{eq:sr1_bloch_gi}).  Using
\beq
\label{eq:vel}
\v=-\frac{i}{\hbar}[\hat\r,\hat H]
\eeq
and  specializing to the $z$-component
of Eq.~(\ref{eq:sr1_wann}),
\beq
\label{eq:sronez}
\sronez=\frac{e}{\hbar cV_c}{\rm Im}\,{\rm tr}_c[\P\x\Q\H\Q\y-\P\H\P\x\Q\y].
\eeq
The second term above may be expanded as a trace in the Wannier
representation as
\beq
\label{eq:PHPxQY}
{\rm tr}_c[\P\H\P\x\Q\y]=\sum_\R\sum_{mn}
\langle\0 m|\H|\R n\rangle\langle \R n|\x\Q\y|\0 m\rangle.
\eeq
Then using the identities
\beq
\label{eq:H_wann}
\langle \0 m|\H|\R n\rangle=V_c\int \frac{d\k}{(2\pi)^3}\,e^{-i\k\cdot\R}
E_{mn\k},
\eeq
\beq
\label{eq:xQy_wann}
\langle \R n|\x\Q\y|\0 m\rangle=V_c\int \frac{d\k}{(2\pi)^3}\,e^{i\k\cdot\R}
\langle\ppartial_x u_{n\k}|\ppartial_y u_{m\k}\rangle,
\eeq
we obtain
\beq
\label{eq:equality1}
\frac{1}{V_c}{\rm tr}_c[\P\H\P\x\Q\y]=
\int \frac{d\k}{(2\pi)^3}\,h_{\k,\alpha\beta}.
\eeq
Using a similar argument, it follows that
\beq
\label{eq:equality2}
\frac{1}{V_c}{\rm tr}_c[\P\x\Q\H\Q\y]=
\int \frac{d\k}{(2\pi)^3}\,g_{\k,\alpha\beta}.
\eeq
Combining the above two equations with Eq.~(\ref{eq:sronez}) then
yields Eq.~(\ref{eq:sr1_bloch_gi}).

\subsection{Gauge-invariant remainder $\deltaM$}
\label{sec:gi_remainder}

To take the thermodynamic limit of $\deltaM$
we start from Eq.~(\ref{eq:deltaM})
and apply it to a large crystallite to arrive at
Eq.~(\ref{eq:deltaM_bloch_gi}). Focusing on the $z$-component,
\beq
(\Delta M)_z=\fac{\rm Tr}[\P\x\Q\hat v_y]-\fac{\rm Tr}[\P\y\Q\hat v_x].
\eeq
Now use Eq.~(\ref{eq:vel}) to obtain
\beq
(\Delta M)_z=-2i\facbar{\rm Tr}[\x\P\y\P\H\P-\y\P\x\P\H\P],
\eeq
where we defined $\facbar=\fac/\hbar$ and replaced $\P\H$ by the more
symmetrical form $\P\H\P$. Using
$-i{\rm Tr}[\hat{\cal O}-{\cal O}^\dagger]=2{\rm Im}\,{\rm Tr}[\hat{\cal O}]$,
this becomes
\beq
\label{eq:deltam_z}
(\Delta M)_z=4\facbar{\rm Im}\,{\rm Tr}[\x\P\y(\P\H\P)].
\eeq

At this point we are still considering a bounded sample.
To obtain a bulk expression we first need to manipulate
Eq.~(\ref{eq:deltam_z}) into a form where the unbounded operators $\x$ and
$\y$ are sandwiched between $\P$ and $\Q$, as in Eq.~(\ref{eq:sronez}). That
ensures that ill-defined diagonal position matrix elements between the
extended Bloch states do not occur. We will make use of the following rules for
finite-dimensional Hermitian matrices $A$, $B$, $C$, and $D$:
\beq
\mbox{(i)}\,{\rm Im}\,{\rm Tr}[ABCD]={\rm Im}\,{\rm Tr}[DABC]
\eeq
\beq
\mbox{(ii)}\,{\rm Im}\,{\rm Tr}[ABCD]=-{\rm Im}\,{\rm Tr}[DCBA]
\eeq
\beq
\label{eq:rule3}
\mbox{(iii)}\,{\rm Im}\,{\rm Tr}[AB]=0
\eeq
and, if any two of the  matrices $A$, $B$, and $C$ commute,
\beq
\mbox{(iv)}\,{\rm Im}\,{\rm Tr}[ABC]=0.
\eeq
Rules (i) and (ii) result from elementary properties of the trace.
Rule (iii) is a consequence of (i) and (ii), and rule (iv) follows from (iii).
Replacing the first $\P$ in Eq.~(\ref{eq:deltam_z}) by $\one-\Q$ and
applying rule (iv) to the term containing
$\one$ ($[\x,\y]=0$ and $\P\H\P$ is Hermitian), we obtain
\beq
\label{eq:deltam_z_b}
(\Delta M)_z=-4\facbar{\rm Im}\,{\rm Tr}[\P\H\P\x\Q\y],
\eeq
which has the desired form.

Now we invoke Wannier-representability to write
\beq
{\rm Tr}[\P\H\P\x\Q\y]=\sum_j\,
\langle w_j|\H\P(\x-\overline{x}_j)\Q(\y-\overline{y}_j)|w_j\rangle
\eeq
(note that $\P\overline\r\Q=0$).
Since only the relative coordinate appears, the
contribution from the surface orbitals is non-extensive, vanishing in the
thermodynamic limit. We are then left with a bulk-like expression:
\beq
{\rm Tr}[\P\H\P\x\Q\y]\rightarrow
\sum_{\R m}\sum_{\R' n}\,
\langle \R m|\H|\R'n\rangle\langle \R' n|\x\Q\y|\R m\rangle.
\eeq
Both matrix elements on the right-hand-side
depend on $\R$ and $\R'$ only through $\R'-\R$, and
therefore, comparing with Eq.~(\ref{eq:PHPxQY}),
\beq
\label{eq:delta_m_wann}
\frac{1}{N_c}{\rm Tr}[\P\H\P\x\Q\y]\rightarrow{\rm tr}_c[\P\H\P\x\Q\y],
\eeq
where $N_c$ is the number of crystalline cells in the sample.
Combining Eqs.~(\ref{eq:equality1}), (\ref{eq:deltam_z_b}) and
(\ref{eq:delta_m_wann}) one obtains Eq.~(\ref{eq:deltaM_bloch_gi}), which
concludes the proof.

\bibliographystyle{apsrev}
\bibliography{./bib}

\begin{thebibliography}{41}
\bibitem[{exp({\natexlab{c}})}]{explan-offdiag}
\bibinfo{note}{However, since the off-diagonal $\r_{ij}$ enter the
  self-rotation as a first power and the spread as a second power, the
  non-invariant part may play a larger role in the former quantity than in the
  latter.}
\bibitem[{exp({\natexlab{d}})}]{explan-intra}
\bibinfo{note}{For metals the low-frequency part of the absorption spectrum
  contains intraband (Drude) contributions in addition to the interband
  absorption. Only the latter was taken into account in the derivation of the
  dichroic $f$-sum rule for crystals.}
\bibitem[{exp({\natexlab{e}})}]{explan-quantum}
\bibinfo{note}{The quantum of polarization can ultimately be traced to the
  quantization of the charge in a Wannier function. On the contrary, neither
  the self-rotation nor the center-of-mass velocity of a Wannier function is
  quantized; these can change continuously under a multiband gauge
  transformation.}
\bibitem[{exp({\natexlab{f}})}]{explan-vanishing}
\bibinfo{note}{To see that the middle expression in Eq.~(\ref{eq:kuhn})
  vanishes, write ${\rm Im}\,{\rm Tr}[\P\hat r_\alpha\Q\hat r_\beta]$ as ${\rm
  Im}\,{\rm Tr}[(\P)(\hat r_\alpha\hat r_\beta)]- {\rm Im}\,{\rm Tr}[(\P\hat
  r_\alpha\P)(\P\hat r_\beta\P)]$, note that $\P$, $\hat r_\alpha\hat r_\beta$,
  $\P\hat r_\alpha\P$ and $\P\hat r_\beta\P$ are all Hermitian, and use
  Eq.~(\ref{eq:rule3}).}
\end{thebibliography}

\end{document}